\documentclass[superscriptaddress,twocolumn]{revtex4}
\usepackage{amsmath,lscape,epsfig}

\def\ii{\'{\i}}
\def\beq{\begin{equation}}
\def\eeq{\end{equation}}
\def\beqa{\begin{eqnarray}}
\def\eeqa{\end{eqnarray}}
\def\ban{\begin{eqnarray*}}
\def\ean{\end{eqnarray*}}
\def\bi{\begin{itemize}}
\def\ei{\end{itemize}}

\def\d{\mbox{d}}

\begin{document}

\title{Constraining relativistic models through heavy ion collisions}

\author{D.P.Menezes}
\affiliation{Depto de F\'{\i}sica - CFM - Universidade Federal de Santa
Catarina  Florian\'opolis - SC - CP. 476 - CEP 88.040 - 900 - Brazil}
\author{C. Provid\^encia}
\affiliation{Centro de F\ii sica Te\'orica - Dep. de F\ii sica -
Universidade de Coimbra - P-3004 - 516 - Coimbra - Portugal}
\author{M. Chiapparini}
\affiliation{Instituto de F\ii sica - Universidade do Estado do Rio de Janeiro
  - Rua S\~ao Francisco Xavier, 524 - CEP 20550-900 - RJ - Brazil}
\author{M.E. Bracco}
\affiliation{Instituto de F\ii sica - Universidade do Estado do Rio de Janeiro
  - Rua S\~ao Francisco Xavier, 524 - CEP 20550-900 - RJ - Brazil}
\author{A. Delfino}
\affiliation{Depto de F\ii sica - Universidade Federal Fluminense -
  Niter\'oi - CEP 24210-150 - RJ - Brazil}
\author{M. Malheiro}
\affiliation{Depto de F\ii sica - Universidade Federal Fluminense -
  Niter\'oi - CEP 24210-150 - RJ - Brazil}
\affiliation{Depto de F\ii sica - Instituto Tecnol\'ogico de
Aeron\'autica, CTA - CEP 12228-900 S\~ao Jos\'e dos Campos - SP -
Brazil}

\begin{abstract}
Relativistic models can be successfully applied to the description
of compact star properties in nuclear astrophysics as well as to
nuclear matter and finite nuclei properties, these studies taking place 
at low and moderate temperatures. Nevertheless, all results are model
dependent and so far it is unclear whether some of them should be
discarded. Moreover, in the regime of hot hadronic matter very
few calculations exist using these relativistic models, in
particular when applied to particle yields in heavy ion collisions.
 In the present work we comment on the known constraints that can help the
 selection of adequate models in this regime
 and investigate the main differences that arise
when the particle production during a Au+Au collision at RHIC is
calculated with different models.
\end{abstract}

\maketitle

\vspace{0.50cm}
PACS number(s): {21.65.+f, 24.10.Jv, 95.30.Tg}
\vspace{0.50cm}

\section{Introduction}

Relativistic models have been widely used in order to describe
nuclear matter, finite nuclei and stellar matter properties. Many
variations of the well known quantum hadrodynamic model \cite{sw}
have been developed and used along the last decades. Some of them
rely on density dependent couplings between the baryons and the
mesons \cite{original,tw,br,gaitanos,twring} while others use
constant couplings \cite{nl3,tm1,glen}. Still another possibility of
including density dependence on the lagrangian density is through
derivative couplings among mesons and baryons
\cite{delf1,delf2,chiappa1} or the coupling of the mediator mesons
among themselves \cite{nlwr}.

All these quantum hadronic models have under control the fitting of the
infinite nuclear matter binding energy at the experimental saturation density
($\rho_0$). Their results are around
  $16MeV$ and $0.15fm^{-3}$, respectively. However at the level of
nuclear matter, they predict different values for other physical quantites, as for instante, incompressibility ($K$), effective
nucleon mass, scalar and vector potentials,
critical temperature ($T_c$), etc.
If applied for finite nuclei, these models also present discrepancies on  nuclear spectra. The efforts to find some
correlations among  observables of infinite nuclear matter and
finite nuclear results, lead to interesting conjectures. One is that there is a correlation between the effective nucleon mass
($M^{*}$) at the saturation nuclear matter density and the spin-orbit
splitting for several nuclei \cite{furn}, for different quantum
hadronic models. This correlation states that  good theoretical L-S splittings
for several finite nuclei are obtained  if $M^{*}$ lies between 0.58
and 0.62. Other correlation was also proposed to connect $T_c$
with $K, M^{*}$  and $\rho_0$ \cite{natowitz}.

There is a strong correlation between compact star properties and
some of the nuclei properties. Relativistic model couplings  are
adjusted in order to fit expected nuclei properties such as binding
energy, saturation density, compressibility and energy symmetry at
saturation density, particle energy levels, etc. Once the same
relativistic model is extrapolated to higher densities as in stellar
matter or higher temperatures as in heavy-ion collisions or even to
lower densities as in the nuclear matter liquid gas phase
transitions \cite{malheiro}, they can and indeed provide
different information. Hence, experimental contraints obtained
either from polarized electron scattered from a heavy target, from
heavy-ion collisions at different energies or from astronomical
observations are very important in order that adequate models are
chosen and inadequated ones are ruled out. In what follows we
discuss some of the already existing constraints and the relation
between equations of state (EoS) used to describe stellar matter and
hadronic matter.

The relation between neutron star properties which are obtained from
specific EoS and the neutron skin thickness has long been
a topic of investigation in the literature. The difference between the neutron
and the proton radii, the neutron skin thickness, is linearly correlated with
the pressure of neutron matter at sub-nuclear densities. This is so because the
properties of neutron stars are obtained from appropriate EoS
(very isospin asymmetric due to the  $\beta$- equilibrium constraint) whose
symmetry energy depends on the density and
also controls the size of the neutron skin thickness in heavy and
asymmetric nuclei, as $^{208}$ Pb, for instance.
In \cite{hp01} it was shown that the models
that yield smaller neutron skins in heavy nuclei tend to yield smaller neutron
star radii due to a softer EoS.
In \cite{peles} it was shown that the neutron skin thickness indeed give hints
on the equations of state that are suitable to describe neutron
stars. A correlation between the slope of the symmetry energy and the neutron
skin thickness, previously found for Skyrme-type models \cite{bao-an}, was
also observed
within relativistic models \cite{peles}. Unfortunately a precise measurement
of the neutron skin thickness is still under way.

Neutron stars are believed to have a solid crust formed by nonuniform neutron
rich matter in $\beta$-equilibrium above a liquid mantle. In the inner crust
nuclei coexist with a gas of neutrons which have dripped out. The properties
of this crust as, for instance, its thickness and pressure at the crust-core
interface depend a lot on the density dependence of the EoS used to describe
it \cite{haen00}. On the other hand, it is well known \cite{chom04,
inst04} that the existence of
phase transitions from liquid to gas phases in asymmetric nuclear matter (ANM)
is intrinsically related with the instability regions which are limited by
the spinodals. Liquid-gas phase transition in ANM can lead to an isospin
distillation phenomenon,  characterized by a larger proton fraction in the
liquid phase than in the gas phase. In \cite{inst062} the spinodal sections
for two relativistic models, one with constant couplings known as NL3
\cite{nl3} and another with density dependent couplings we usually refer to as
TW, were obtained.
The curve that represent the proton versus neutron densities for
$\beta$-equilibrium matter was represented in the same plot and one can see
that it always crosses the spinodal section at $T=0$. This means that a
liquid-gas phase transition occurs at the crust of a cold neutron star
giving  rise to a nonhomogeneous region. For higher temperatures this effect
generally does not occur meaning that the outer layer of a hot compact star is
homogeneous. For $T=10$ MeV, for instance, the two models investigated gave
contradictory information as for the NL3 the crossing was observed and for the
TW it was not. More information is then required so that the correct picture
can be depicted and the wrong EoS ruled out.

\begin{figure}[h]
\includegraphics[width=8.cm]{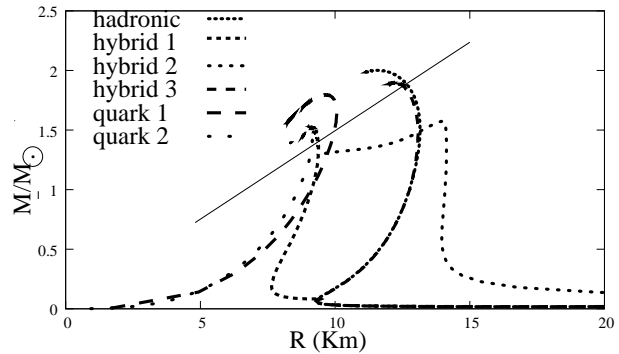}
\caption{Mass-radius plots for one hadronic, three hybrid and two quark stars.}
\label{figtov}
\end{figure}

Compact star properties can be computed from the solution of the
Tolman-Oppenheimer-Volkoff (TOV) equations \cite{tov} once they are supposed
to be spherically symmetric and static. The input to the TOV equations is the
EoS chosen to describe the stellar matter. The
determination of neutron star properties obtained from the  measureament of
the gravitational redshift of spectral lines produced in neutron star
photosphere provides a direct constraint on the mass-to-radius ratio.
Some time ago a redshift of 0.35 from three different transitions of the
spectra of the X-ray binary EXO0748-676 was obtained in \cite{cottam}.
This redshift corresponds to $M/R=0.15 M_\odot/Km$. The 1E 1207.4-5209 neutron
star, which is in the center of the supernova remnant PKS 1209-51/52 was also
observed and two absorption features in the source spectrum were detected
\cite{sanwal}. These features were associated
with atomic transitions of once-ionized helium in the neutron star atmosphere
with a strong magnetic field. This interpretation leads to a readshift of the
order of 0.12-0.23, considerably lower than the one in \cite{cottam}. This
readshift imposes another constraint to the mass to radius ratio given by
$M/R=0.069 M_\odot/Km$ to $M/R=0.115 M_\odot/Km$.
However, the interpretation of the absorption features as atomic transition
lines in \cite{sanwal} is controversial: an alternative interpretation
\cite{bignami,xu03} is that the absorption features are cyclotron lines,
which imply no obvious constraint on the EoS. In previous works
\cite{deltas,kaons,quark1} many EoS were tested against these two possible
constraints. While all EoS studied so far are consistent with the measurement
proposed by \cite{sanwal}, many of them are excluded if the redshift found in
\cite{cottam} provides the only possible constraint. As an example we plot
the mass and radii obtained from different EoS in Fig. \ref{figtov}, where we
have added the line corresponding to the more restrictive constraint
(top straight line).

The hadronic star \cite{hadronic} was built with the inclusion of 8 baryons
within the non-linear Walecka model and parametrization GM1 \cite{glen} given
in \ref{tab1}.
The hybrid stars were obtained with the same model and parametrization as
above for the hadron phase and different models for the quark phase. Hybrid 1
\cite{bursts} mixed phase contained the MIT bag model \cite{mit} with
$Bag=(160{\,\rm MeV})^4$, hybrid 2 \cite{bursts} the color flavored
locked phase (CFL) \cite{cfl} again with $Bag=(160{\,\rm MeV})^4$
and hybrid 3 \cite{hybrid3} the Namb-Jona-Lasinio (NJL) model \cite{njl}.
The quark star properties were obtained for the CFL with the same bag
parameter as above and called quark 1 and with the NJL model and called quark 2
\cite{bursts}. One can notice that while hybrid 2 star is excluded, hybrid 1
and hybrid 3 stars remain valid. The hadronic and quark stars remain as
possible candidates.

Electron-positron pairs can be emitted from bare quark stars \cite{pairs}
and this dominant emission requires the existence of a surface layer of
electrons tied to the star by a strong electric field. While the electron
chemical potential of a quark star described by the MIT bag model is very low
(less than 20 MeV), the NJL model gives much higher values reaching 100 MeV
inside the star \cite{quark2}.

A very recent paper \cite{rikovska} reported a series of possible
constraints used separately to analyse relativistic and
non-relativistic equations of state applied to neutron stars. The
authors mention a relation derived in \cite{podsia} that correlates
gravitational and baryonic neutron star masses and the measurement
of Pulsar A moment of inertia \cite{lattimer} which excludes some of
the standard mean field relativistic models.

\noindent From the above considerations, one can see that different
relativistic models in the description of compact stars provide
different information. In order to identify the possible
constituents of compact objects and, in this way, rule out some of
the relativistic models, more astronomical observations are
necessary.

Another attempt in the direction of choosing appropriate models is
discussed in the present work in the high temperature regime produced in 
heavy ion collisions. For this reason we
consider Au-Au collisions at RHIC/BNL and analyse the hadron
abundances and particle ratios in order to determine the temperature
and baryonic chemical potential of the possibly present phase
transition. The possibility that a quark-gluon plasma (QGP) could be
formed in heavy ion collisions arose when quantum chromodynamics
(QCD) at finite temperature and high densities became a topic of
increasing interest due to the discovery of asymptotic freedom about
30 years ago. In cosmology, the relevant conditions for QGP
formation occur $10\rm\,\mu$s after the Big-Bang theory; nuclear
matter first appears after about $1\rm\,ms$. In laboratory searches
for QGP, in large  colliders around the world (RHIC/BNL, ALICE/CERN,
GSI, etc), experimentalists are trying to do the opposite: to
convert hadronic matter at sufficiently high temperatures into QGP.

A first step towards the interpretation of data collected during a
Pb+Pb collision at the SPS was taken in \cite{munzinger}, where the
authors have used a statistical model which assumes chemical
equilibration to find the temperature and baryon chemical potential
that provide a best fit to the data obtained by the NA49 \cite{na49}
and WA97 \cite{wa97} colaborations. They concluded that the chemical
freezeout temperature was $T=168 \pm 2.4$ MeV and the corresponding
baryon chemical potential was $\mu_B=266 \pm 5$ MeV. Lattice Monte
Carlo simulations of QCD at vanishing baryon density gives a very
similar estimate, i.e., $T=170 \pm 8$ MeV \cite{lattice}. Latter on,
the same model was used to analyse the particle production yields
measured in central Au+Au collision at RHIC \cite{munzinger2} and
the authors obtained $T=174 \pm 7$ MeV and $\mu_B=46 \pm 5$ MeV. In
both cases \cite{munzinger,munzinger2} they used the free Fermi
and Boson gas approximations where the interaction among the baryons
and mesons were neglected and there were only the above mentioned
independent parameters, i.e., the temperature and baryon chemical
potential. Three conservation laws were imposed: of the baryon
number, the strangeness and the electric charge. While the
temperatures found were of the same order, the very different
chemical potentials imply different baryon density. In a more recent
paper \cite{schaffner}, a simple relativistic selfconsistent chiral
model, where the nuclear interaction was included, was
employed to analyse the Au+Au collision at RHIC once again. The
results depend on the parametrizations used but they found a
lower temperature around $T=155$ MeV and the baryon chemical
potential of the order of $\mu_B=51$ MeV. The authors claim that the
fitted chemical freeze-out temperatures and chemical potentials
depend on the order of the phase transition and suggest that at RHIC
the system emerges after the phase transition.

In the present work we revisit the same data and analyse them with the help of
more sofisticated relativistic models within a mean field approach. We use
four parametrizations of the non-linear Walecka model \cite{sw}, namely
NL3 \cite{nl3}, TM1 \cite{tm1}, GM1 and GM3 \cite{glen}, one model with
implicit density dependence through meson field couplings, the
NL$\omega\rho$ \cite{nlwr} and two different parametrizations of a
density dependent hadronic model, the TW \cite{tw} and the DDME1
\cite{twring}. Similar investigations were performed in \cite{chiappa} within
one of the parametrizations mentioned below (GM1) and the best fit resulted in
$T=143.9$ MeV and $\mu_B=25.07$ MeV. We need to understand how model dependent
these numbers are. {Hence}, for comparison,  we also investigate the results obtained within a
model containing  the eight lightest baryons (baryonic octet) without any
interaction and then with the baryonic octet plus extra ten baryons
(baryonic decuplet) also
without interaction to try to understand how important the interaction at low
densities are.
We implemented a fit based on the minimum value of the
quadratic deviation as in \cite{munzinger} in
order to obtain the temperature and chemical potential for each model.
The paper is organized as follows: in section II we show the lagrangian
densities of the models considered and describe the formalism used; in section
III we present and discuss the results; in section IV we draw our final
conclusions.

\section{Formalism}

In the following we review the basic formulae of the models we use. Just the
Lagrangian density and the main expressions for densities and chemical
potentials are given. If one wants to follow
the analytical calculations in detail, please refer to \cite{sw,peles,ddpeos,
supernova}, among many other papers in the literature.

\subsection{Different parametrizations for the non-linear Walecka model}

The  Lagrangian density that incorporates many parametrizations of the
non-linear Walecka model (NLWM) \cite{sw,nl3,tm1,glen} and also the extra
non-linear
$\sigma-\rho$ and $\omega-\rho$ couplings \cite{nlwr} reads

$$
{\cal L}=\sum_B \bar \psi_B \left[\gamma_\mu \left( i\partial^{\mu}- g_{vB}
V^{\mu}- g_{\rho B} {\boldsymbol {\tau}} \cdot \mathbf {b}^\mu \right. \right.
$$
$$ \left. \left. -e  \frac{(1+2 \tau_{3B})}{2} A^\mu \right)
-(M-g_{s B} \phi)\right]\psi_B$$
$$ +\frac{1}{2}(\partial_{\mu}\phi\partial^{\mu}\phi
-m_s^2 \phi^2)
-\frac{1}{3!}\kappa \phi ^{3}
-\frac{1}{4!}\lambda \phi ^{4}$$
$$-\frac{1}{4} \Omega_{\mu\nu}\Omega^{\mu\nu}
+\frac{1}{2} m_v^2 V_{\mu}V^{\mu}
+\frac{1}{4!} \xi gv^4 (V_\mu V^\mu)^2 $$
$$-\frac{1}{4}\mathbf B_{\mu\nu}\cdot\mathbf B^{\mu\nu}
+\frac{1}{2} m_\rho^2 \mathbf b_{\mu}\cdot \mathbf b^{\mu}
-\frac{1}{4}F_{\mu\nu}F^{\mu\nu}
$$
\begin{equation}
+g_{\rho B}^2\mathbf b_{\mu}\cdot \mathbf b^{\mu}[\Lambda_s
g_{s B}^2\phi^2+\Lambda_v g_{v B}^2 V_\mu V^\mu],
\label{lagacoplada}
\end{equation}
where $B$ represents the lightest { 18} baryons,
$\phi$, $V^\mu$, $\mathbf {b}^\mu$ and $A^{\mu}$ are the
scalar-isoscalar, vector-isoscalar and vector-isovector meson
fields and the photon field respectively,
$\Omega_{\mu\nu}=\partial_{\mu}V_{\nu}-\partial_{\nu}V_{\mu}$ ,
$\mathbf B_{\mu\nu}=\partial_{\mu}\mathbf b_{\nu}-\partial_{\nu} \mathbf b_{\mu}
- \Gamma_\rho (\mathbf b_\mu \times \mathbf b_\nu)$,
$F_{\mu\nu}=\partial_{\mu}A_{\nu}-\partial_{\nu}A_{\mu}$ and
$\tau_{3B}$ is the third component of the baryon isospin.
The  parameters of the model are:
the nucleon mass, of the order of $M=939$ MeV, depending on the
parametrization used, the masses of
the mesons $m_s$, $m_v$, $m_\rho$, also model dependent,
the electromagnetic coupling constant
$e=\sqrt{4\pi/137}$ and the coupling constants between baryons and mesons
$$g_{s B}=x_{s B}~ g_s,~~g_{v B}=x_{v B}~ g_v,~~g_{\rho B}=x_{\rho B}~
g_{\rho},$$
where $x_{s B}$, $x_{v B}$ and $x_{\rho B}$ are equal to $1$ for the nucleons
{ and deltas} and acquire different values in different parametrizations
for the other baryons. { In the present work we have used
$x_{s B}=0.7$, $x_{v B}=0.783$ and $x_{\rho B}=0.783$ for couplings between
mesons and $\Lambda, \Sigma, \Xi, \Sigma^*, \Xi^*$ and
$x_{s B}=x_{v B}=x_{\rho B}=0$ between mesons and the $\Omega$.}

Non-linear $\sigma$ terms are also included in all parametrizations through
the constants $\kappa$ and $\lambda$ and a non-linear $\omega$ term is present
in the TM1 \cite{tm1} parametrization through the constant $\xi$.
A density dependence is introduced through the non-linear
$\sigma-\rho$ and $\omega-\rho$ couplings \cite{nlwr}, just present in the
$NL\omega\rho$ model.
We have followed the prescription of \cite{hp01}, where the starting point was
the NL3 parametrization and the $g_\rho$ coupling was adjusted for each value
of the coupling $\Lambda_i$ studied in such a way that for
$k_F=1.15$ fm$^{-1}$ (not the saturation point) the symmetry energy is
25.68 MeV. In the present work we set $\Lambda_s=0$ as in \cite{nlwr}.
Other possibilities for this model with
$\sigma-\rho$ and $\omega-\rho$ couplings have already been discussed in the
literature. All coupling constants are adjusted in order to reproduce the
nuclear matter saturation properties given in Table \ref{tab1}.

\subsection{Density dependent hadronic model}

The Lagrangian density of the density dependent model we use next reads:
$$
{\cal L}=\sum_B \bar \psi_B \left[\gamma_\mu\left(i\partial^{\mu}-\Gamma_{v B}
V^{\mu}- \Gamma_{\rho B}  \boldsymbol{\tau} \cdot \mathbf {b}^\mu
\right. \right.
$$
$$ \left. \left. -e  \frac{(1+ 2 \tau_{3B})}{2} A^\mu \right)
-(M-\Gamma_{s B} \phi)\right]\psi_B
$$
$$
+\frac{1}{2}(\partial_{\mu}\phi\partial^{\mu}\phi
-m_s^2 \phi^2)
-\frac{1}{4}\Omega_{\mu\nu}\Omega^{\mu\nu}$$
\begin{equation}
+\frac{1}{2}m_v^2 V_{\mu}V^{\mu}
-\frac{1}{4}\mathbf B_{\mu\nu}\cdot\mathbf B^{\mu\nu}+\frac{1}{2}
m_\rho^2 \mathbf b_{\mu}\cdot \mathbf b^{\mu}
-\frac{1}{4}F_{\mu\nu}F^{\mu\nu}
\label{lagtw}
\end{equation}
where
$\Omega_{\mu\nu}$, $\mathbf B_{\mu\nu}$ and $F_{\mu\nu}$ are defined after
eq.(\ref{lagacoplada}). The  parameters of the model are again the masses and
the couplings, which are
now density dependent, i.e., $\Gamma_s$ replaces $g_s$, $\Gamma_v$ replaces
$g_v$ and $\Gamma_\rho$ replaces $g_\rho$. { Once again the relations
$$\Gamma_{s B}=x_{s B}~ \Gamma_s,~~\Gamma_{v B}=x_{v B}~
\Gamma_v,~~\Gamma_{\rho B}=x_{\rho B}~ \Gamma_{\rho}$$
hold and}
these density dependent couplings $\Gamma_{s}$,
$\Gamma_v$ and $\Gamma_{\rho}$ are adjusted in order to reproduce
some of the nuclear matter bulk properties shown in Table \ref{tab1},
using the following parametrization:
\begin{equation}
\Gamma _{i}(\rho )=\Gamma _{i}(\rho _{sat})h_{i}(x),\quad x=\rho /\rho _{sat},
\label{paratw1}
\end{equation}
with
\begin{equation}
h_{i}(x)=a_{i}\frac{1+b_{i}(x+d_{i})^{2}}{1+c_{i}(x+d_{i})^{2}},\quad i=s,v
\end{equation}
and
\begin{equation}
h_{\rho }(x)=\exp [-a_{\rho }(x-1)],  \label{paratw2}
\end{equation}
with the values of the parameters $m_{i}$,
$\Gamma _{i}(\rho_{sat})$, $a_{i}$, $b_{i}$, $c_{i}$ and $d_{i}$,
$i=s,v,\rho $ given in \cite{tw}.
This model does not include self-interaction terms for the meson
fields (i.e. $\kappa =0$, $\lambda =0$ and $\xi=0$ ) as in NL3 or TM1
parametrizations for the NLWM.

{Two parametrizations of this density dependent hadronic models are
  discussed next. The original one, that we refer to as TW and a more recent
parametrization known as DDME1 \cite{twring}, also shown to provide
a good description for the properties of many stable nuclei.}

\subsection{Calculations}

We start by applying the Euler-Lagrange equations to the Lagrangian densities
and obtaining the mesonic equations of motion. Then a mean field approximation
is enforced and the equations of motion, which have to be solved in a
self-consistent way become dependent of the baryonic densities. At this point
the electromagnetic field is neglected. For the
different parametrizations of the non-linear Walecka model, the densities read:

\begin{equation}
\rho_B=(2J_B+1) \int\frac{\d^3p}{(2\pi)^3}(f_{B+}-f_{B-}), \quad
\rho=\sum_B \rho_B,
\label{rhob}
\end{equation}
with $M_B^*=M_B - g_{s B}~ \phi$, $B\pm$ stands respectively for baryons
and anti-baryons,
$E^{\ast}({\mathbf p})=\sqrt{{\mathbf p}^2+{M^*}^2}$ and
\begin{equation}
f_{B\pm}=
{1}/\{1+\exp[(E^{\ast}({\mathbf p}) \mp \nu_B)/T]\}\;,
\end{equation}
where the effective chemical potential is
\begin{equation}
\nu_B=\mu_B - g_{v B} V_0 - g_{\rho B}~  \tau_{3 B}~ b_0.
\end{equation}
The baryon chemical potentials are computed in terms of their quark
constituints, i.e., $\mu_n= \mu_u + 2 \mu_d$ and related expressions for the
others. As particles and anti-particles have to be computed separately for the
calculation of their yields, after the self-consistent calculation they are
calculated as
\begin{equation}
\rho_{B+}= (2J_B+1) \int\frac{\d^3p}{(2\pi)^3}f_{B+},
\label{rhob+}
\end{equation}
\begin{equation}
\rho_{B-}=(2J_B+1)  \int\frac{\d^3p}{(2\pi)^3}f_{B-}.
\label{rhob_}
\end{equation}
{where $J_B=1/2$ and 3/2 respectively for the baryonic octet and decuplet.}
For the density dependent hadronic model, the expressions are very similar
except for the
effective mass $M_B^*=M_B - \Gamma_{s B}~ \phi$ and
the effective chemical potential
\begin{equation}
\nu_B= \mu_B-\Gamma_{v B} V_0 - \Gamma_{\rho B}~ \tau_{3B}~ b_0
- {\Sigma^{R}_0}_{TW},
\end{equation}
where the rearrangement term ${\Sigma^{R}_0}_{TW}$ is given by
$$ {\Sigma^{R}_0}_{TW}=\sum_B \left[
\frac{\partial \Gamma_{v B}}{\partial \rho} \rho_B V_0 +
\frac{\partial \Gamma_{\rho B}}{\partial \rho} \tau_{3b}~ \rho_B~ b_0 \right.$$
\begin{equation}
\left. -\frac{\partial \Gamma_{s B}}{\partial \rho} \rho_{s B} \phi_0,
\right].
\label{rearrtw}
\end{equation}

Moreover, as we are interested in obtaining also the production of pions and
kaons, they are introduced through Bose-Einstein distribution functions
\begin{equation}
\rho_i =
\frac{2J_M+1}{2\pi^2} \int_0^\infty p^2 dp
\left[\frac{1}{exp[(E_i-\mu_i)/T] -1} \right],
\label{rhopi}
\end{equation}
{ where $i=\pi^+,\pi^-,\pi^0, K^+, K^-,K^0, \bar K^0$, and the corresponding
vector mesons $\rho$ and $K^*$, with $J_M=0$ and 1.
$E_i=\sqrt{p^2 + m_i^2}$ and the chemical potencials are again written in
terms of their quark constituints, namely,
$\mu_{\pi^+}=\mu_u-\mu_d$, $mu_{K^+}=\mu_u-\mu_s$ and so on. We have
considered that they behave like a free gas and their properties are not
changed due to their interaction with matter and, therefore, the fraction of produced
mesons is determined statisitically from their free space properties.  }

In order to obtain the particle yields and respective densities three
conserved quantities are considered: the total strangeness is set to zero, the
total number of baryons in a Au+Au collision is $N_B= 2 (N+Z)= 394$ and the
total isospin is $I_3=(Z-N)/2=-39$. Our code deals with 6 unknows, the three
meson
fields and the three independent quark chemical potentials ($\mu_q, q=u,d,s$),
solved in a self-consistent manner.

Finally, as a last test, we have disregarded the interactions by setting all
couplings equal to zero. We have also performed the same test, i.e., for free
Fermi particles and  including the ten baryons with total spin 3/2,
as suggested in \cite{schaffner}.
We next analyse our results.

\section{Results}

As already mentioned, in obtaining our results we have fixed the
relations between the baryons and the mesons as
$x_{s B} = 0.7$, $x_{v B} = 0.783 $ and $x_{\rho B} = 0.783$, { except for
  the nucleons and deltas, when they are 1 and for the $\Omega$'s, when they
  are zero.} We have verified that other choices give slightly different
results but have decided to fix just one possibility to restrict our test to
the model differences.

As stated in the Introduction,
we have  implemented a $\chi^2$ fit as in \cite{munzinger} in
order to obtain the temperature and chemical potential for each model.
In Table \ref{tab2} we show our results for the different models studied
  corresponding to the temperature and chemical potential that for each model
  give the minimum value of the quadratic deviation   $\chi^2$:

\begin{equation}
\chi^2 = \sum_i \frac{({\cal R}_i^{exp} -{\cal R}_i^{theo})^2}
{\sigma_i^2},
\end{equation}
and
\begin{equation}
q^2 = \sum_i \frac{({\cal R}_i^{exp} -{\cal R}_i^{theo})^2}
{({\cal R}_i^{theo})^2},
\end{equation}
where ${\cal R}_i^{exp}$ and ${\cal R}_i^{theo}$ are the $i^{th}$ particle
ratio given experimentally and calculated with our models and $\sigma_i$
represents the errors in the experimental data points.

In obtaining the best fit values for the temperature and chemical potentials,
we have used the experimental ratios appearing in Table \ref{tab2} four times
for $\bar p/p$, twice for $\pi^-/\pi^+$ and four times for $K^-/K^+$, all with
the same weight. We have also taken into account the $K^{0*}/h^-$ and
$\bar K^{0*}/h^-$ ratios, where $h^-$ is the net sum of all negative
electrically charged hadrons. Instead we could have taken the mean value of
  the measured values and a statistical average value of the errors. We have
  checked that the results were similar.

 One can observe from Table \ref{tab2} that we have found incredibly good
  agreements between experimental data and all tested models. The only
  exceptions are the free octet and free octet+decuplet models that cannot
  describe the $\bar p/\pi^-$ ratio. Although the inclusion of the decuplet
improves the result, it remains far behind the experimental value.
These results show that a proper treatment of the interaction has to be
considered. In \cite{munzinger}, to account for the repulsive interaction
between particles an eigenvolume was assigned to all particles.
We have obtained
approximately the same freeze-out temperature for all models,
namely, $\simeq 146.5-152.5$ MeV. The higher limit is close to the value
obtained with two parametrizations in \cite{schaffner}. As for the chemical
potentials, our results are very model dependent. While all four
parametrizatins of the NLWM $\mu_B$
predict a similar chemical potential $\sim 47-48$ MeV, equal or slighlty lower
than the prediction  of \cite{schaffner}, where the
encountered value lies between 48.3 and 54.6 MeV, the density dependent models
predict larger values, respectively 59 and 62.5 for
  DDME1 and TW.
To test the importance of including the interaction between particles we have
set the couplings to zero. We have obtained  a worse  $\chi^2$ value compared with the
values obtained when interactions are included. Although we get a similar
freeze-out temperature, the chemical potentials obtained with no interation
are 15-20 MeV smaller.

In Table II we include other ratios not yet measured. We conclude that they are
not very sentive to the model except $\bar \Omega/\Omega$.  The results that
are worse reproduced involve mesons, $K-/\pi-$. This seems to indicate that our
treatment of the mesons, with constituent masses equal to the vacuum values, is
too naive and an improvement should be considered. We have checked that, if we
renormalize the mass of the mesons according to the results given in
\cite{schaffner} namely, consider the pion and kaon mass 1.1 times the vacuum
mass and the $K^*$ mass 0.9 of the vaccuum mass we would get
$\chi^2=5.3$ and $T=149.5$ MeV,  $\mu_B= 48.75$MeV for the NL3 parametrization.

For the sake of completeness, in Table \ref{tab3} we display a sample of
hadron effective masses. In this table $N$ stands for nucleons. One can
  see that, with very few exceptions, they are basically model independent and
  the influence of the medium is rather low for all baryons, excluding the
  nucleons. For nucleons the effective mass is at most 15\% smaller than the
  vacuum values. For the other hyperons we get 5-6\% reduction. These masses
  seem to show that the freeze out occurs below the critical temperature. This
means that the freeze-out occurs at a temperature below the phase transition
to a massless baryon phase.

\section{Conclusions}

In the present work we discussed the necessity of finding
constraints able to discard or select appropriate relativistic models.
As all models are parametrized to fix nuclear matter saturation properties,
once they are extrapolated to lower or higher densities or finite temperature,
all sorts of results can turn up. Concerning astronomical measurements,
observations that could help in ruling out possible mass-radius relations
are most welcome. Constraints can also come from the measurement of a precise
neutron skin thickness or from heavy ion collisions, the topic of our
calculations here.

By analysing the particle densities and production yields obtained in a Au+Au
collision at RHIC we have verified that all models describe the data in a
similar way and therefore these data can not be used to establish a constraint.
This is due to the fact that the chemical
potential and consequently, the baryonic density involved in the hadronization
process are very low.
We have also confirmed that in order to improve the fit we must go beyond the
the naive way mesons were included.

 The analysis of the results of the Pb+Pb
collision at SPS may be more adequate in the search for constraints to
different models since, at least, with thermal models, the chemical potential
seems to be much larger. This work is under investigation and, if proved to
be an adequate choice, some improvements can be done
for the achievement of a better description. The inclusion of
strange mesons as mediators of the hyperons \cite{strange}, responsible for
explaining the strongly attractive hyperon-hyperon interaction as observed in
double $\Lambda$ hypernuclei will certainly influence the densities of
particles containing strangeness. Also, the inclusion of
the scalar isovector virtual $\delta (a_0(980))$ field, that introduces
in the isovector channel the structure of relativistic
interactions and affects the behavior of the system at high densities or high
temperatures \cite{deltas} can also bring some modifications. Still other
relativistic models as the quark-meson-coupling model (QMC) \cite{qmc} that
presents a quarkionic structure inside the hadrons should be investigated.
These calculations are already under way.
On the other hand, more data is also required.

\section*{ACKNOWLEDGMENTS}

This work was partially supported by CNPq(Brazil) and FEDER/FCT (Portugal)
under the projects  POCI/FP/63918/2005 and PDCT/FP/63912/2005.

\newpage

\begin{table}
\caption{ Nuclear matter properties.}
\begin{center}
\begin{tabular}{cccccccccccc}
\hline
ratio &  NL3 & TM1 & GM1 & GM3  & & & NL$\omega\rho$ & & & TW & DDME1 \\
& \cite{nl3} &\cite{tm1} & \cite{glen} & \cite{glen} & && \cite{nlwr} & &
&\cite{tw} & \cite{twring}\\
\hline
& & & & & & $\Lambda_v=0.01$ &  $\Lambda_v=0.02$ &  $\Lambda_v=0.025$ & & &\\
\hline
$B/A$ (MeV) & 16.3 & 16.3 & 16.3 & 16.3 & & 16.3 & 16.3 & 16.3 && 16.3 & 16.2 \\
$\rho_0$ (fm$^{-3}$) & 0.148 & 0.145 & 0.153 & 0.153 && 0.148 & 0.148 & 0.148
&& 0.153 & 0.152 \\
$K$ (MeV) & 271 & 281 & 300 & 240 && 271   & 271 & 271 && 240 & 244.5\\
${\cal E}_{sym.}$ (MeV) & 37.4 & 36.9 & 32.5 & 32.5 && 34.9 & 33.1 & 32.3 &&
32.0 & 33.1 \\
$M^*/M$ & 0.60 & 0.63 & 0.70 & 0.78 && 0.60 & 0.60 & 0.60 && 0.56 & 0.578\\
\hline
\end{tabular}
\end{center}
\label{tab1}
\end{table}

\newpage

\begin{table}
\caption{ Comparison of experimental particle ratios and relativistic mean
field models.}
\begin{center}
\begin{tabular}{cccccccccccc}
\hline
ratio & exp. data & exp &  NL3 & TM1 & GM1 & GM3  & TW & DDME1 & octet & octet +\\
&&& \cite{nl3} &\cite{tm1} & \cite{glen} & \cite{glen} & \cite{tw} & \cite{twring} && decuplet  \\
\hline
$\bar{p}/p$  & 0.65$\pm$0.07 & STAR & 0.650 & 0.646 & 0.626 & 0.597 & 0.656 &
0.663 & 0.661 & 0.649 \\
             & 0.64$\pm$0.07 & PHENIX & & & & & & & & &\\
             & 0.60$\pm$0.07 & PHOBOS & & & & & & & & &\\
             & 0.64$\pm$0.07 & BRAHMS & & & & & & & & &\\
$\bar{p}/\pi^-$ & 0.08$\pm$0.01 & STAR & 0.075 & 0.072 & 0.072 & 0.063 & 0.076
& 0.074 & 0.039 & 0.041 \\
$\pi^-/\pi^+$   & 1.00$\pm$0.02 & PHOBOS & 0.998 & 0.991 & 0.999 & 1.00 & 1.01
& 1.01 & 1.00 & 1.01 \\

                & 0.95$\pm$0.06 & BRAHMS & & & & & & &&&\\
$K^-/K^+$       & 0.88$\pm$0.05 & STAR & 0.912 & 0.911 & 0.907 & 0.905 & 0.896
& 0.900 & 0.961 & 0.941 \\
                & 0.78$\pm$0.13 & PHENIX && && &&&&&\\
                & 0.91$\pm$0.09 & PHOBOS && && &&&&&\\
                & 0.89$\pm$0.07 & BRAHMS && && &&&&&\\
$K^-/\pi^-$     & 0.149$\pm$0.02& STAR & 0.234 & 0.234 & 0.242 & 0.243 & 0.228
& 0.227 & 0.232 & 0.235 \\
$\bar{\Lambda}/\Lambda$ & 0.77$\pm$0.07 & STAR & 0.681 & 0.680 & 0.666 & 0.644
&0.663 & 0.675 & 0.687 & 0.689 \\
$\bar \Xi^-/\Xi^-$ & 0.82$\pm$0.08 & STAR & 0.746 & 0.747 & 0.735 & 0.711 &
0.739 & 0.748 & 0.714 & 0.732 \\
$K^{0*}/h^-$ & 0.06 $\pm$ 0.017 & STAR & 0.058 & 0.059 & 0.063 & 0.064 & 0.064 & 0.063 & 0.060 & 0.061 \\
$\bar K^{0*}/h^-$ &0.058 $\pm$ 0.017 & STAR & 0.053 & 0.054 & 0.057 & 0.058 &
0.056 & 0.056 & 0.057 & 0.057 \\
\hline
\hline
$\bar \Omega/\Omega$ &&& 0.693 & 0.699 & 0.715 & 0.723 & 0.585 & 0.586 & - & 0.784\\
$\bar \Omega/\pi^-$ &&& 0.001 & 0.001 & 0.002 & 0.002 & 0.001 & 0.001 & - & 0.001 \\

$\Lambda/h^-$ &&& 0.021 & 0.020 & 0.021 & 0.020 & 0.023 & 0.022 & 0.013 & 0.014 \\
$\Omega/\Xi^-$ &&& 0.172 & 0.178 & 0.195 & 0.211 & 0.173 & 0.174 & - & 0.250 \\
$\lambda/K^{0*}$ &&& 0.351 & 0.341 & 0.331 & 0.301 & 0.363 & 0.353 & 0.226 & 0.228\\
$\bar \Xi^-/\Lambda$ &&&0.226 & 0.226 & 0.227 & 0.221 & 0.296 & 0.295 & 0.241 &
0.222 \\
$\bar \Xi^-/\bar \Lambda$ &&&0.332 & 0.332 & 0.341 & 0.343 & 0.330 & 0.327 &
0.312 & 0.322 \\
$\Xi^-/\bar K^-$ &&& 0.047 & 0.045 & 0.050 & 0.049 & 0.048 & 0.045 & 0.027 &
0.030\\
$\bar \Xi^-/\bar K^-$ &&& 0.035 & 0.034 & 0.035 & 0.033 & - & - & 0.019 & 0.022 \\
\hline
$T$(MeV) &&& 149 & 149 & 152. & 152.8 & 146.6 & 146.2 & 146.4 & 148.8 \\
$\mu_b$ (MeV) &&& 47.5 & 46.5 & 47.5 & 48. & 62.8 & 57. & 30.5 & 32.5 \\
$\rho$ $\times 10^{-3}$ (fm$^{-3}$) &&& 8.37 & 8.03 & 9.62 & 9.95 & 4.90 & 4.45
& 2.41 & 4.77 \\
$\chi^2$ &&& 23.94 & 24.43 & 27.99 & 33.19 & 22.18 & 21.83 & 45.44 & 41.63 \\
radius (fm) &&& 22.4 & 22.7 & 21.38 & 21.14 & 26.77 & 27.65 & 33.91 & 27.02 \\
\hline
\end{tabular}
\end{center}
\label{tab2}
\end{table}

\begin{table}
\caption{Effective masses obtained with the temperature and chemical
  potentials given in Table \ref{tab2}.}
\begin{center}
\begin{tabular}{cccccccccccc}
\hline
&  NL3 & TM1 & GM1 & GM3  & TW & DDME1 \\
& \cite{nl3} &\cite{tm1} & \cite{glen} & \cite{glen} & \cite{tw} & \cite{twring} \\
\hline
$M^*_N/M_N$ & 0.88 & 0.89 & 0.90 & 0.93 & 0.87 & 0.87 \\
$M^*_\Lambda/M_\Lambda$ & 0.93 & 0.93 & 0.94 & 0.96 & 0.92 & 0.92 \\
$M^*_\Sigma/M_\Sigma$ & 0.94 & 0.94 & 0.95 & 0.96 & 0.93 & 0.93 \\
$M^*_\Xi/M_\Xi$ & 0.94 & 0.95 & 0.95 & 0.96 & 0.93 & 0.94 \\
$M^*_\Delta/M_\Delta$ & 0.91 & 0.92 & 0.93 & 0.95 & 0.90 & 0.90 \\
$M^*_{\Sigma^*}/M_{\Sigma^*}$ & 0.94 & 0.95 & 0.95 & 0.95 & 0.94 & 0.94 \\
$M^*_{\Xi^*}/M_{\Xi^*}$ & 0.95 & 0.95 & 0.96 & 0.97 & 0.94 & 0.94 \\
$M^*_\Omega/M_\Omega$ & 1.00 & 1.00 & 1.00 & 1.00 & 1.00 & 1.00 \\
\hline
\end{tabular}
\end{center}
\label{tab3}
\end{table}

\end{document}